\documentclass[prd,letterpaper,superscriptaddress,reprint,nofootinbib]{elsarticle} 

\usepackage{floatrow}
\usepackage{amssymb}
\usepackage{amsthm}
\usepackage{amsmath}
\usepackage{amsbsy}
\usepackage{nicefrac}
\usepackage{hyperref}
\usepackage{wrapfig}
\usepackage{lineno}
\usepackage{graphicx}
\include{rlFig}

\begin{document}

\begin{frontmatter}
\title{Experimental tests of sub-surface reflectors as an explanation for the ANITA anomalous events}
\author[Chicago]{D.~Smith}
\author[KU,MEPHI]{D.~Z.~Besson}
\author[Chicago]{C.~Deaconu}
\author[OSU,CCAP]{S.~Prohira}
\author[OSU,CCAP]{P.~Allison}
\author[UCL]{L.~Batten}
\author[OSU]{J.~J.~Beatty}

\author[WashU]{W.~R.~Binns}
\author[WashU]{V.~Bugaev}
\author[UD]{P.~Cao}
\author[NTU]{C.~Chen}
\author[NTU]{P.~Chen}
\author[UD]{J.~M.~Clem}
\author[OSU,CCAP]{A.~Connolly}
\author[UCL]{L.~Cremonesi}
\author[IITT]{P.~Dasgupta}
\author[UH]{P.~W.~Gorham}
\author[WashU]{M.~H.~Israel}
\author[NTU]{T.~C.~Liu}
\author[UCLA]{A.~Ludwig}
\author[UH]{S.~Matsuno}
\author[UH]{C.~Miki}
\author[NTU]{J.~Nam}
\author[KU]{A.~Novikov}
\author[UCL]{R.~J.~Nichol}
\author[Chicago]{E.~Oberla}
\author[UH]{R.~Prechelt}
\author[WashU]{B.~F.~Rauch}
\author[UH]{J.~Russell}
\author[UCLA]{D.~Saltzberg}
\author[UD]{D.~Seckel}
\author[UH]{G.~S.~Varner}
\author[Chicago]{A.~G.~Vieregg}
\author[PSU]{S.~A.~Wissel}

\address[UCLA]{Dept. of Physics and Astronomy, Univ. of California, Los Angeles, Los Angeles, CA 90095.}
\address[OSU]{Dept. of Physics, Ohio State Univ., Columbus, OH 43210.}
\address[UH]{Dept. of Physics and Astronomy, Univ. of Hawaii, Manoa, HI 96822.}
\address[NTU]{Dept. of Physics, Grad. Inst. of Astrophys.,\& Leung Center for 
Cosmology and Particle Astrophysics, National Taiwan University, Taipei, Taiwan.}
\address[UCI]{Dept. of Physics, Univ. of California, Irvine, CA 92697.}
\address[KU]{Dept. of Physics and Astronomy, Univ. of Kansas, Lawrence, KS 66045.}
\address[WashU]{Dept. of Physics and McDonnell Center for the Space Sciences, Washington Univ. in St. Louis, MO 63130.}
\address[SLAC]{SLAC National Accelerator Laboratory, Menlo Park, CA, 94025.}
\address[UD]{Dept. of Physics, Univ. of Delaware, Newark, DE 19716.}
\address[UCL]{Dept. of Physics and Astronomy, University College London, London, United Kingdom.}
\address[UMinn]{School of Physics and Astronomy, Univ. of Minnesota, Minneapolis, MN 55455.}
\address[CCAP]{Center for Cosmology and Particle Astrophysics, Ohio State Univ., Columbus, OH 43210.}
\address[Chicago]{Dept. of Physics, Enrico Fermi Institute, Kavli Institute for Cosmological Physics, Univ. of Chicago , Chicago IL 60637.}
\address[PSU]{Dept. of Physics, The Pennsylvania State University, State College, PA 16801.}
\address[CalPoly]{Dept. of Physics, California Polytechnic State Univ., San Luis Obispo, CA 93407.}
\address[IITT]{Dept. of Physics, Indian Institute of Technology, Kanpur, Uttar Pradesh 208016, India}
\address[MEPHI]{National Research Nuclear University, Moscow Engineering Physics Institute, 31 Kashirskoye Highway, Rossia 115409}


\date{\today}
\begin{abstract}
The balloon-borne ANITA\cite{GorhamAllisonBarwick2009} experiment is designed to detect ultra-high energy neutrinos via
radio emissions produced by an in-ice shower. Although initially purposed for
interactions within the Antarctic ice sheet, ANITA also demonstrated the ability to self-trigger on radio emissions from ultra-high energy charged cosmic rays\cite{HooverNamGorham2010} interacting in the Earth's atmosphere. 
For showers produced above the Antarctic ice sheet,
reflection of the down-coming radio signals at the Antarctic surface should result in a 
polarity inversion prior to subsequent observation at the $\sim$35-40 km altitude ANITA gondola. 
ANITA has published two anomalous instances of upcoming cosmic-rays 
with measured polarity opposite the remaining
sample of $\sim$50 UHECR  signals\cite{Gorham:2016zah,gorham2018observation}.
The steep observed upwards incidence angles (25--30 degrees relative to the horizontal) require non-Standard Model physics if these events are due to in-ice neutrino interactions, as the Standard Model cross-section would otherwise prohibit neutrinos
from penetrating the long required chord of Earth.
Shoemaker {\it et al.}\cite{shoemaker2019reflections} posit that glaciological effects
may explain the steep observed anomalous events.
We herein consider the scenarios offered by Shoemaker {\it et al.} and find them to be disfavored by extant ANITA and HiCal experimental data.  We note that the recent report of four additional near-horizon anomalous ANITA-4 events\cite{ANITA4ME}, at $>3\sigma$ significance, are incompatible with their model, which requires significant signal transmission into the ice. 

\end{abstract}
\end{frontmatter}

\section*{Introduction}
The origins of, and acceleration mechanisms responsible for the highest-energy cosmic rays observed at Earth are currently
a source of considerable speculation. Given their low interaction likelihood and inertness to galactic and inter-galactic
magnetic fields, observation of ultra-high energy (UHE) neutrinos ($E >$ 1 EeV) is of particular interest, as they may reveal cosmic
accelerators otherwise obscured by the Cosmic Microwave Background (CMB).
Several experiments have been commissioned within the last 20 years with the goal of the first-ever detection of 
UHE neutrinos originating from beyond the 
Milky Way. 
 The small neutrino flux at such energies requires
large target volumes. Cold Antarctic ice, with radio-frequency attenuation lengths exceeding 1 km\cite{Barwick2005south} is
therefore attractive as an experimental neutrino target due to the broadband radio-frequency Askaryan emission expected to be produced when UHE neutrinos interact in the ice. 
Attempts are being made to instrument the ice with radio receivers to detect this emission, either with near-surface antennas~\cite{Barwick:2014rca}, antennas drilled deeper into the ice\cite{allison2012design}, or from a high-elevation synoptic platform as with ANITA~\cite{GorhamAllisonBarwick2009}. ANITA's balloon-borne receivers have many orders of magnitude more ice in the field of view, and therefore provide the greatest geometric acceptance, but at a cost of a higher threshold due to the greater typical distance to candidate events and challenges in characterizing backgrounds due to the changing field of view. 

While designed to look primarily for the Askaryan emission from in-ice UHE neutrino interactions, another type of physics signature is accessible to these broadband radio receivers. 
Down-coming UHE cosmic-ray (UHECR) hadrons colliding with atmospheric molecules produce extensive air showers. At shower maximum, the relative ratio of $\gamma:e^\pm:\mu^\pm$:hadrons $\approx$ 3000:300:3:1, with a slight excess of $e^-$ over $e^+$ owing to the Askaryan effect. The $e^-/e^+$ number asymmetry and the Lorentz force of the local geomagnetic field on the charged constituents of the shower produce significant radio-frequency (RF) power in the band from 30--1000 MHz\cite{timZ2017ptep}, with a net electric field polarization resulting from a superposition of the Askaryan Cherenkov cone (${\vec E}$ radially outwards along the conical Cherenkov front) with the unidirectional field resulting from the Lorentz force. Since 80\%-90\% of the signal is due to the geomagnetic contribution, the polarity is primarily determined by the local ${\vec v}\times{\vec B}$.
Those radio signals are typically confined to a hollowed
Cerenkov-like cone approximately $\theta_C\sim 0.7^o$ in half-width
projecting to a two-dimensional annulus in the transverse plane; the
annulus has transverse thickness $\delta\theta\sim\theta_C/2\sim 0.3^o$\cite{Schoorlemmer:2015afa}
 following a lateral
Gaussian signal strength profile centered at $\theta_C$.

The ANITA experiment demonstrated the viability of the synoptic strategy with the first self-triggered observation of radio 
emissions from ultra-high energy cosmic rays interacting in the atmosphere\cite{HooverNamGorham2010}. 
Three flights yielded a total of approximately 50
such detected events 
(the sensitivity of the ANITA-II flight was reduced due to a trigger sensitive only to vertically-polarized signals)
consistent with the impulsive characteristics expected for radio emissions from cosmic rays\cite{gorham2018observation}. 
Two events coming from the ice exhibited polarity opposite that
seen in reflected UHECR and were interpreted as upcoming and thus
`anomalous' events (AE).
 ANITA showed that those events exhibit signal attributes consistent
with upcoming air showers. Such showers may be generated by the atmospheric decay of $\tau$ leptons produced in charged-current interactions of $\tau$-neutrinos in the Earth, although that hypothesis requires suppressing the Standard Model neutrino
cross-section, 
extrapolated from existing data up to the EeV scale. 
Analysis by ANITA\cite{romero2019comprehensive} as well as a  recent analysis by the IceCube\cite{aartsen2020search} collaboration experimentally disfavors the tau neutrino hypothesis. 
It has also been suggested that radio emissions from an air shower core impacting the Antarctic surface at steeper incidence angles may explain these events without invoking Beyond Standard Model physics\cite{de2019coherent}. 

AE1 and AE2 are reconstructed at Antarctic surface source locations (latitude, longitude) of (-82.6559 S, 17.2842 E) and (-81.39856 S, 129.01626 E), respectively, and with $\sim$20 km error ellipses projected back to the Antarctic surface.
Observation of four additional candidate events in the ANITA-4 data set has also very recently been reported\cite{StephNeutrino2020Conf}. In contrast to the anomalous polarity events AE1 and AE2, the four additional events are observed just below the horizon. Such a glancing geometry is better matched to the tau hypothesis, although it requires positing a flaring source to evade bounds from the IceCube experiment. In what follows, we focus on AE1 and AE2, although the techniques presented herein can be readily extended to AE3--AE6.

\section*{Glaciological Explanations for the ANITA Events} 
Shoemaker {\it et al.}\cite{shoemaker2019reflections} have proposed that the observed signal polarities are a consequence of simple
glaciology and offer several possible explanations for the origins of the ANITA anomalous events, including
i) multiple layers with $\sim$decimeter spacing and measurable reflectivity per layer, 
ii) firn density inversions, 
iii) surface wind/ablation crusts, sastrugi and/or similar deviations from smoothness, leading to non-specular reflections, 
iv) ice fabric layers,
v) subglacial lakes (below rather than embedded within the ice sheet; Lake Vostok, e.g.), 
vi) snow-covered crevasses, and 
vii) englacial layers. 
To match the existing data, 
the authors require that such features be present over $\sim$7\% of the Antarctic continent, and also that a large-enough fraction of the incident 
signal penetrates the surface so as to produce a triggerable signal at the ANITA payload.

\subsection*{Impact on implied UHECR energy spectrum} Since the sub-surface reflections require transmission of signal into the ice, the implied energy of the UHECR progenitor in the Shoemaker {\it et al.} model must be higher than that of standard surface-reflected signals. Shoemaker {\it et al.} calculate the numerical impact of their model on the detected UHECR energy spectrum in the Appendix of their publication, under the assumption that the observed reflection is due to an extended $\mathcal{O}$(100 m) under-dense cavity in the ice rather than a local density enhancement, such that the signal requires transmission through only one layer to produce the observed anomalous polarity. For an $E^{-2.7}$ primary charged cosmic ray spectrum, we note that an attenuation of 80\% of signal strength (typical of the magnitude required to produce the desired signal polarity from an over-dense sub-surface reflector) corresponds to a 99.6\% reduction in the UHECR flux available to produce the anomalous events.

\subsection*{Models Considered}
Below, we attempt to more quantitatively assess the likelihood that such effects might explain the ANITA observations using existing data supplemented by data from the ANITA and HiCal experiments. 
HiCal\cite{prohira2019hical} was proposed as a high-altitude calibration RF source, emitting narrow pulses as it trailed ANITA by 100--1000 km, allowing quantification of surface reflectivity effects by comparison of HiCal signals observed both directly ({\tt D}) as well as via their surface reflections ({\tt R}) by ANITA\cite{prohira2018antarctic,gorham2017antarctic}. 

Shoemaker {\it et~al.}'s favored model consists of sub-glacial reflectors (lakes, e.g.), and a general class of sub-surface reflectors ({\tt SSR}, including density contrasts, embedded reflectors, and fabric contrasts, e.g., with each layer characterized by a thickness $t$ and separation between layers $d$), and surface effects (sastrugi, e.g.). Before considering sub-surface reflectors, we briefly consider alternative glaciological explanations:
\begin{itemize}
\item Ice-sheet fabrics: 
The macroscopic bulk alignment of ice crystals is described by the so-called `ice fabric', which can be measured via thin-slice analysis of ice core 
samples\cite{Voigt2017}. Discontinuities between ice fabric domains, and/or
realignment of ice fabric within an ice sheet can, in principle, result in weak, but measurable radar echoes\cite{fujita2003scattering}. However,
as discontinuities in ice fabrics typically extend over multi-meter scales with correspondingly `soft' extended multi-meter scale boundaries, and are additionally characterized by very small reflection coefficients (${\cal R}\sim-$60 dB), they would seem unable to produce the sharp, ns-risetime signals observed by ANITA.

\item Wind/ablation crusts and sastrugi: 
Since deviations from ideal specular scattering may result in a loss of signal fidelity across a reflecting surface,
surface roughness effects may be related to the ANITA anomalies. However, surface features
are likely to be time- and location-specific, with strong dependence on the local recent wind history; reconstructing the exact surface topography at the time and location of the ANITA 
anomalies is therefore challenging given the limited continental sampling. A rigorous calculation\cite{dasgupta2018general} shows that extreme surface topographies may yield distortions in reflected waveforms that could potentially result in reflected polarity opposite to that
expected, however, the statistical probability of such features is likely considerably smaller than 7\%, especially considering the high measured signal-to-noise of AE1 and AE2.

\item Sub-glacial lakes:
At the coordinates of AE1 and AE2, the ice depth has
been measured from Ground-Penetrating Radar echo returns to be 3.53 and 3.26 km, respectively.
The quoted elevation angles for AE1 and AE2 (-27.4 and -35.0 degrees, respectively) would correspond
to emergence angles (with respect to vertical) of approximately 
68.7 and 61.1 degrees, respectively, for the two events.
The possibility of sub-glacial lakes, it seems, is therefore ruled out by ice attenuation - 
accounting for the incidence angles of the rays relative to the Antarctic surface, the total in-ice pathlengths for AE1 and AE2
are approximately 9.55 
and 8.14 km, respectively. 
Assuming an attenuation length, averaged over the entire ice sheet equivalent to measurements made at South Pole
($\langle L_{atten} \rangle\sim$700 meters)\cite{barwick2005south} suppresses the signal amplitude by a factor ${\cal O}(10^{-6}-10^{-5})$, 
implying a significant UHECR flux well beyond the GZK-cutoff, and in conflict with existing data.

Sub-glacial lakes are, in fact, excluded on more fundamental grounds. Since the permittivity of water is greater than that of ice, reflection at the ice-water interface under the ice sheet would fail to produce a polarity inversion. Reflection off the bottom of water-bedrock interface could produce non-inversion, but would result in an even more highly-suppressed signal. 

\item Tribo-electric Effect:
Although not included in the Shoemaker {\it et al.} paper, we also consider static discharges in the class of geophysical phenomena that might, in principle, produce spurious broadband backgrounds for ANITA, albeit with no obviously preferred polarization.
In addition to sculpting topography, it has long been realized that wind 
blowing over particulate surface layers may generate significant voltages across the surface, leading to the generation of short-duration, high-amplitude RF via the `tribo-electric effect'\cite{gordon2009electric,schmidt1999electrostatic}, with 
measured electric field strengths as high as 30 kV/m. 
This effect has been observed as correlations 
of local wind velocity $v_{wind}$
with experimental trigger rates, as well as channel-by-channel
root-mean-square voltages in Antarctic experiments, including RICE\cite{KravchenkoFrichterSeckel2003,Kravchenko:2011im}, AURA\cite{Landsman2007aura}, ARIANNA\cite{Barwick:2014rca} and ARA\cite{allison2012design}. Empirical evidence suggests that tribo-electric emissions are measurable provided $v_{wind}$ exceeds some threshold -- taken together, those experiments imply a threshold $v_{wind}\gtrsim$12 m/s\footnote{Additional details on experimental measurements of triboelectric effect in Antarctic experiments will be provided in a forthcoming publication.}.  
Although there are no meteorological stations at the specific locations of AE1 and AE2, there is, nevertheless, wind velocity data available from numerous locations on the Antarctic continent that allow qualitative assessment of this possibility. At the time that 
the AE1 trigger was registered by ANITA-1 (2006-12-28, 00:33:20 UTC), none of the seven
active Antarctic weather stations recorded wind speeds exceeding 12 m/s; 
only weather station Theresa (Latitude 84.60S  Longitude: 115.81W  Elev: 1463m, and therefore 1300 km distant from the reconstructed AE1 source location) registered a wind velocity as high as 11 m/s. At the time the 
AE2 trigger was recorded by ANITA-3 (2014-12-20, 08:33:22.5 UTC), the 13 active weather
stations all recorded wind speeds all below 10 m/s. The nearest weather station with available data (AGO4, 500 km distant), in fact, did not record wind speeds exceeding 10 m/s for the entirety of December, 2014. Four years of wind velocity data for AGO4
are shown in Fig. \ref{fig:AGO4} (left), illustrating the dearth of wind velocities above the nominal tribo-electric threshold of 12 m/s.

\begin{figure}[htpb]
\includegraphics[width=0.45\textwidth]{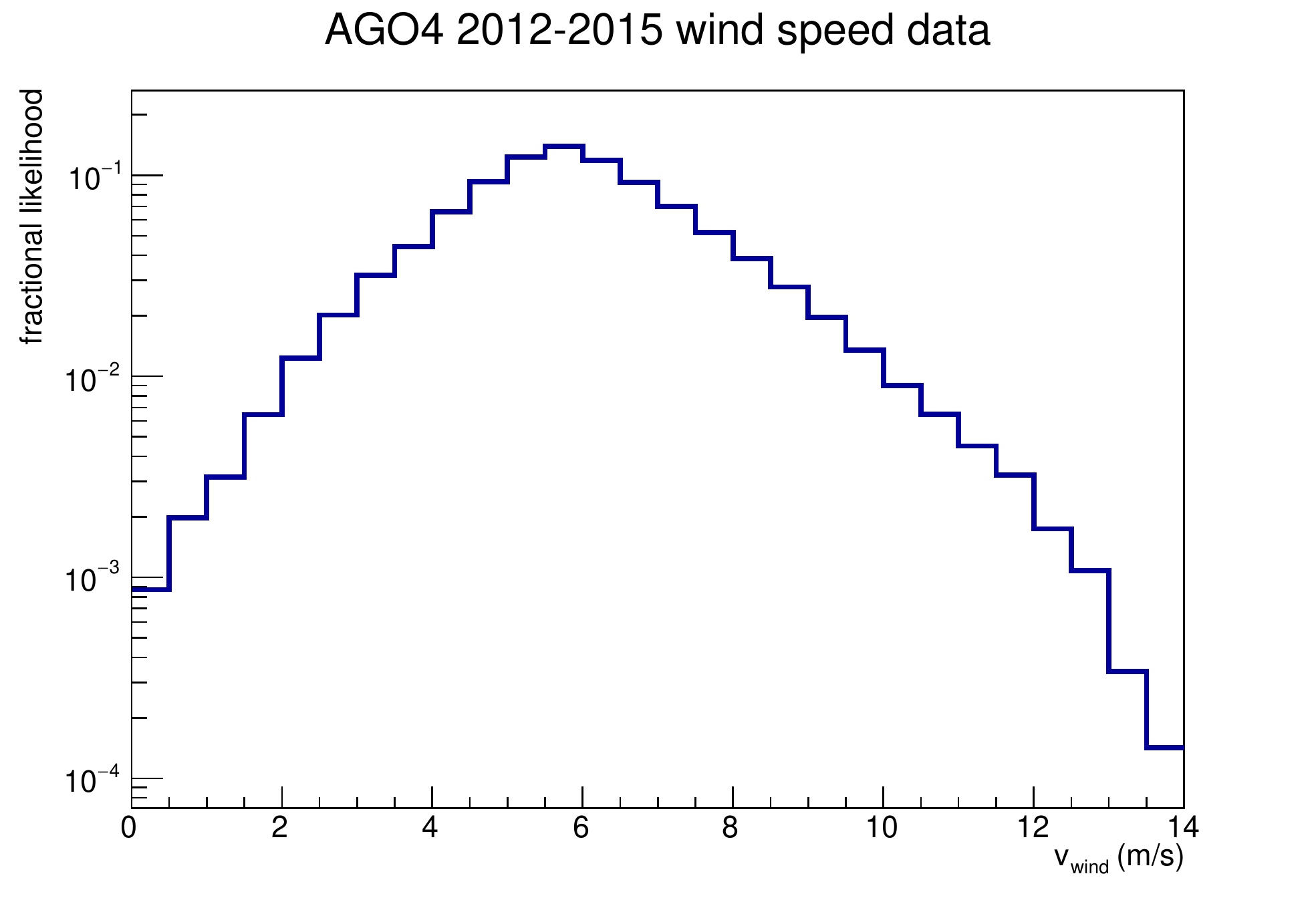} 
\includegraphics[width=0.45\textwidth]{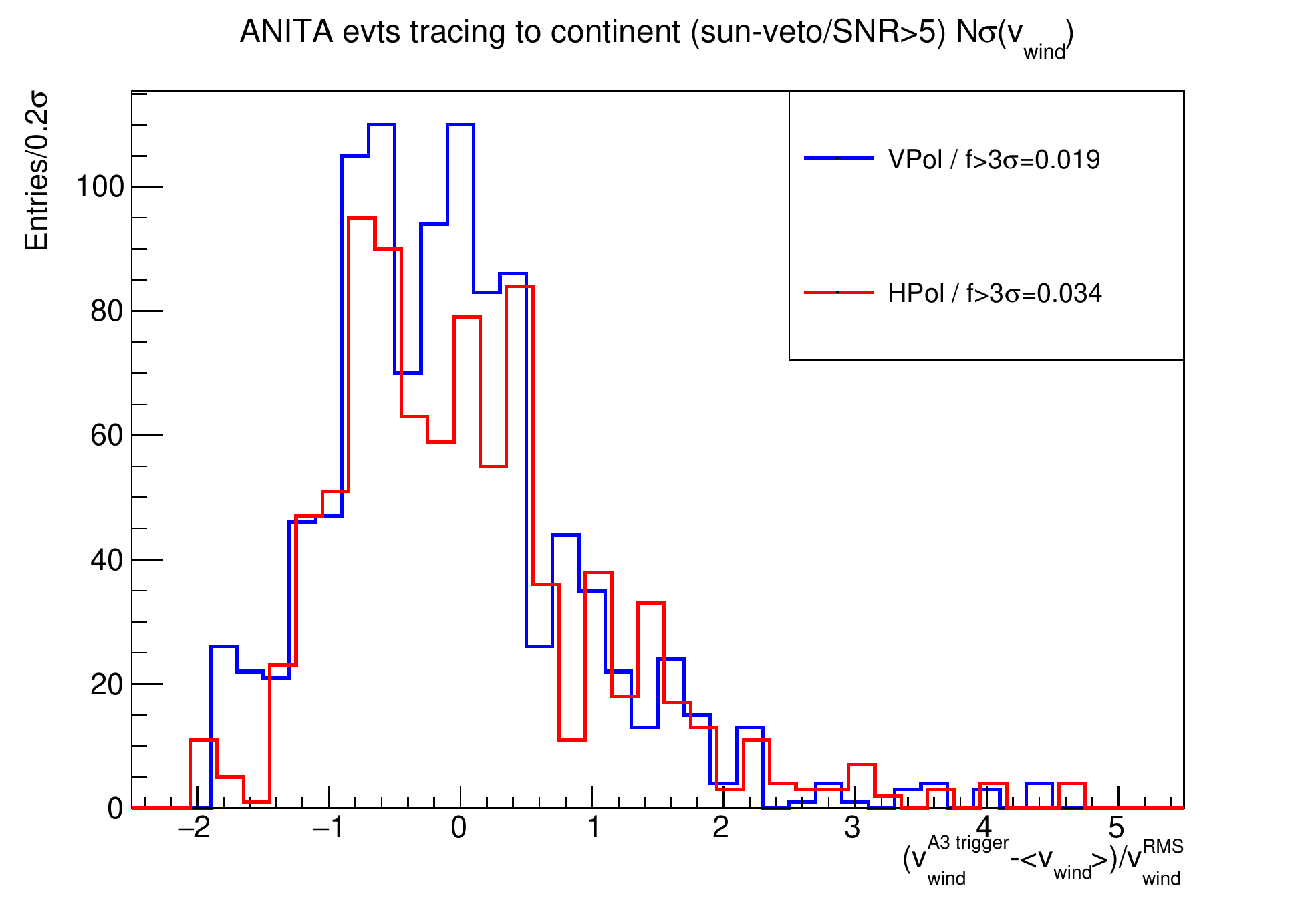}
\caption{(Left:) AGO4 wind velocity data, 2012--2015; (Right:) For that Antarctic weather station closest to the reconstructed source location of an ANITA-3 trigger, deviation between instantaneous wind velocity (at time of recorded trigger) relative to average for that station, in units of rms of wind velocity distribution.}
\label{fig:AGO4}
\end{figure}

Also shown in Fig. \ref{fig:AGO4} (right) is the result of a dedicated search for correlations between wind velocity and ANITA triggers using ANITA-3 data. For this search, we a) determine the average windspeed for each of the Antarctic stations ($\sim$40) for which data was compiled during the ANITA-3 flight, b) for each ANITA-3 trigger with a peak in that event's interferometric map i) having a signal-to-noise ratio exceeding 5.0, ii) not reconstructing to the location of the sun at the time the trigger was recorded, and iii) pointing to a location on the surface of the Antarctic continent, we tabulate the wind velocity at that station closest to the projected source location, and determine the deviation $\delta$ between that wind velocity datum relative to the average for that station. We observe a slight positive tail in the $\delta$ distribution for both VPol and HPol; in all cases for which $\delta>3$, the recorded wind speed exceeds 11 m/s. Nevertheless, we note that: a) since wind velocity data are taken at 10 minute intervals, while ANITA typically triggers at 50 Hz, the distribution shown is non-statistical and has multiple entries per wind velocity datum, and b) as many of the stations that record data are wind-powered through the winter, we have not excluded the possibility that the correlation may be the result of radio-frequency noise produced by wind turbine generators vs. the triboelectric effect, per se.


Global weather models, such as the NOAA's Global Forecast System (GFS)~\cite{GFS} and the the models from the European Centre for Medium-Range Weather Forecasts (ECMWF)\cite{ECMWF} combine data from the world-wide network of  weather stations to produce a global grid of meteorological conditions at regular time intervals, typically for the purpose of weather forecasting. Reanalysis data is also available, applying the latest models to historical data. Among the many parameters computed at each grid point is the the wind velocity at an altitude of 10 m, from which the surface wind speed may be estimated via scaling relations. Historical GFS models show a 10 m elevation wind speed (gusts) of less than 7 m/s (11 m/s)  at the location and time of AE2 and less than 2 m/s (gust data not available) at the location and time of AE1. ECMWF reanalysis data, available at higher resolution, produces consistent results. The wind speed at 10 m is depicted in Fig.~\ref{fig:windmodel}.  In obstruction-less environments such as Antarctica, the surface wind speed is typically 75\% of the the 10 m wind speed, suggesting that surface wind speeds likely did not exceed 8 m/s at the time of these two events, safely below the tribo-electric threshold. 
\end{itemize}

\begin{figure}[htpb] 
\includegraphics[width=3.25in]{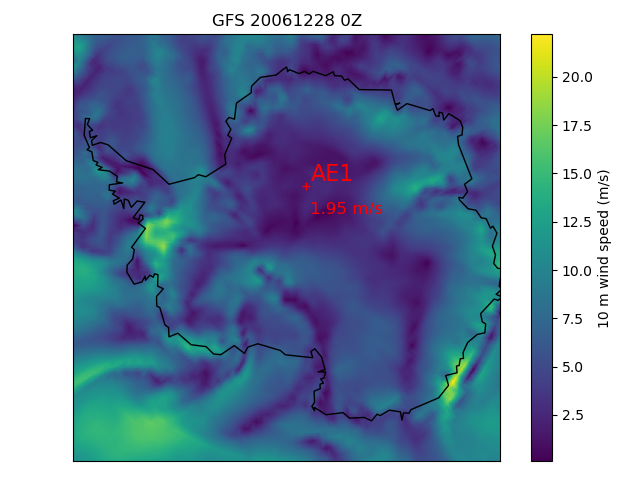} 
\includegraphics[width=3.25in]{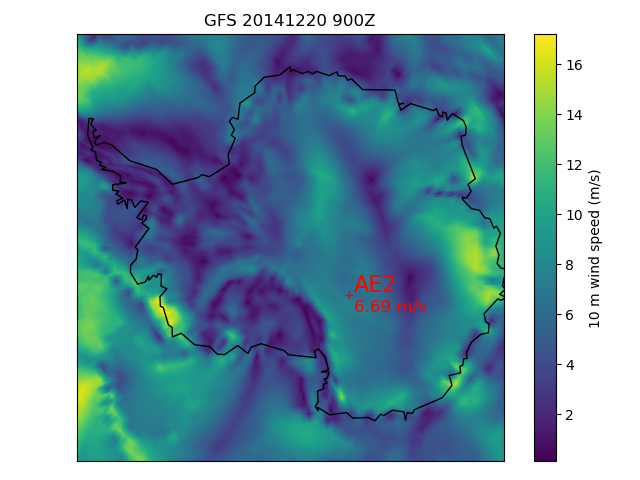} \\
\includegraphics[width=3.25in]{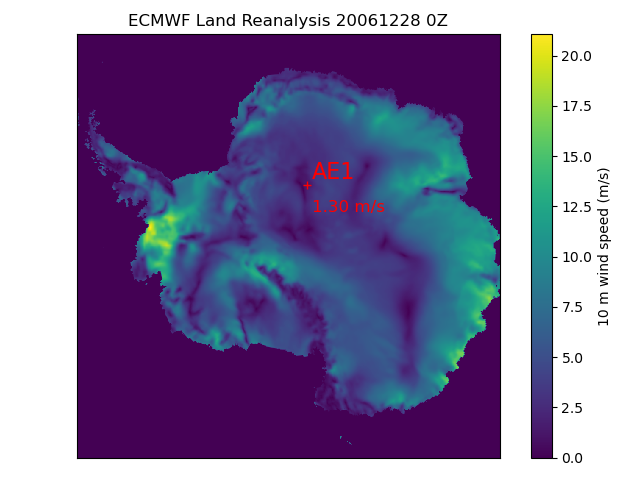} 
\includegraphics[width=3.25in]{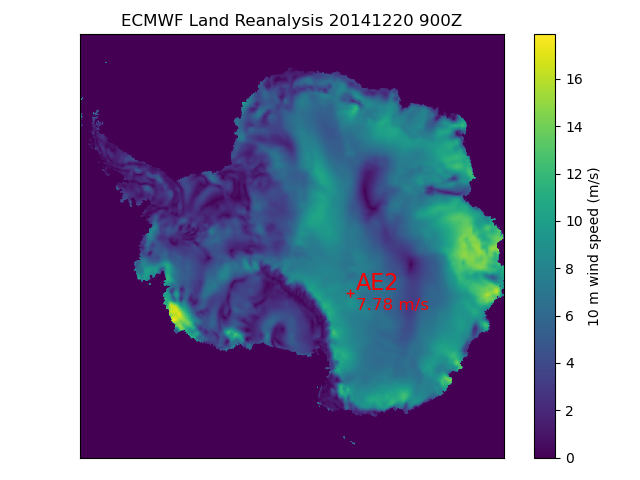} 
\caption{The magnitude of wind speed at a height of 10 m from the closest-in time GFS (top) and ECMWF (bottom, only land shown) models for AE1 and AE2.  The surface wind, which would be responsible for any tribo-electric emission, is generally  75\% of the 10 m wind and likely was not close to the nominal triboelectric threshold wind speed of 12 m/s at the time these events were recorded.}
\label{fig:windmodel} 
\end{figure} 

\section*{Previous measures of reflectivity:}
We next consider the possibility that the ANITA anomalous events originate from a point on the continent characterized by enhanced reflectivity so as to evade the flux suppression arguments above. 
We first review existing data from both satellites and also ANITA, followed by data recorded using HiCal-broadcast signals. In general, presented data address the question of a putative sub-surface reflector only qualitatively and circumstantially.
More quantitative confrontation with the Shoemaker {\it et al.} model follows later in this article.

\subsection*{EviSat Data}
`Radar Altimetry' consists of a radar wave beamed in the nadir direction and received by an on-board sensor after surface reflection.
High-precision satellites (particularly the European Remote Sensing satellites ERS-1\cite{attema1991active} and ERS-2\cite{davis2004elevation}) 
collecting 13.6 GHz Ku-band radar data measure surface elevation by timing radar echoes. 
 The surface height is derived from the travel time and precise knowledge of the satellite location. From waveform information, these satellites 
also quantify position-dependent reflectivity.
The reflected waveform captured by the on-board receiver records the energy initially back-scattered from the surface plus any sub-surface reflectors ({\tt SSR}) at later times. The rising/falling slopes of the radar echoes therefore provide information on both non-specular surface scattering and {\tt SSR} at a given location. Specifically,
the leading edge of the waveform is related to 
the surface roughness and the near-surface characteristics, whereas the trailing edge encodes volume scattering and any non-specular surface features\cite{remy2009antarctic}.

\begin{figure}[htpb]\includegraphics[width=0.75\textwidth]{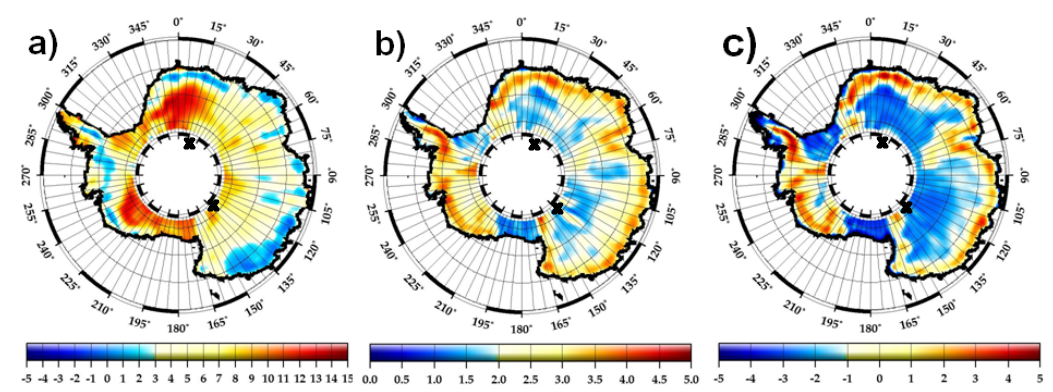} \caption{EviSat data, showing a) Backscattering Coefficient (dB), b) leading edge of backscattered waveform, translated to estimated 
distance (m) scale of possible non-specular surface features, c) trailing edge of backscattered waveform, in units of MHz (adapted from Fig. 4 of \cite{remy2009antarctic}). Overlaid ``X'' shows locations of anomalous ANITA events.}\label{fig:EviSat}\end{figure}

Reflectivity data from the EviSat\cite{remy2009antarctic} satellite are shown in Figure \ref{fig:EviSat} with the locations of AE1 and AE2 overlaid. Anomalous, location-specific sub-surface reflectors would presumably be evidenced by a local `discoloration' at the sites of the mystery event reflections. Although AE1 is beyond the geographical
coverage of EviSat, the location around AE2 does not obviously indicate scattering anomalies.

\subsection*{ANITA-2 solar data}
\begin{figure}[htpb]\centerline{\includegraphics[width=0.8\textwidth]{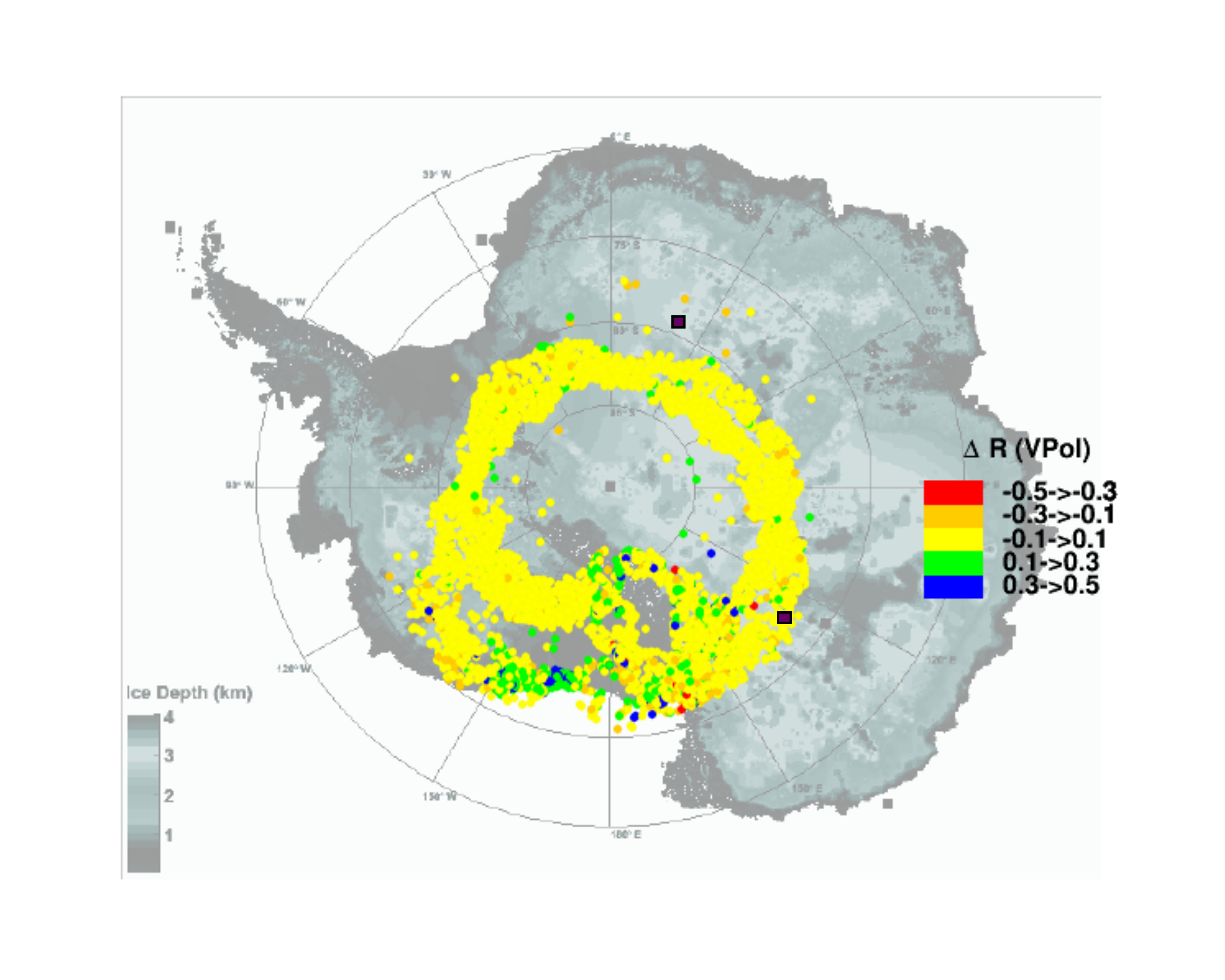}}\caption{Deviation between measured VPol reflectivity and expected (calculated from Fresnel coefficients) for ANITA-2 observation of solar signal, with location of AE1 and AE2 overlaid (black squares).}\label{fig:A2VPolSolarPower}\end{figure}

\begin{figure}[htpb]\centerline{\includegraphics[width=0.8\textwidth]{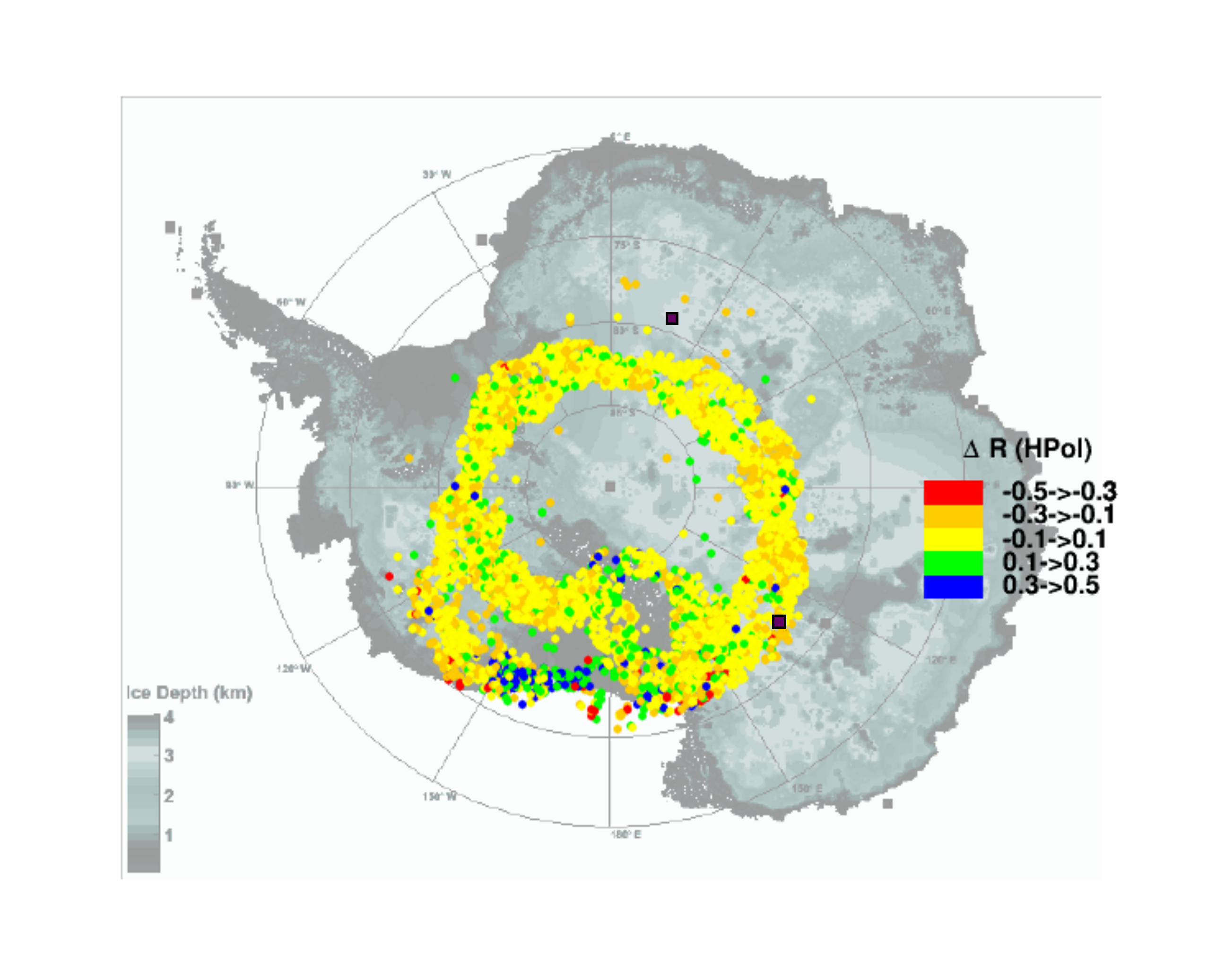}}\caption{Deviation between measured HPol reflectivity and expected (calculated from Fresnel coefficients) for ANITA-2 observation of solar signal, with location of AE1 and AE2 overlaid (black squares).}\label{fig:A2HPolSolarPower}\end{figure}

Owing to its unobscured view, as well as full azimuthal coverage, ANITA has excellent sensitivity to radio-frequency emissions from the Sun, observed via direct ray paths, as well as ray paths reflected from the Antarctic surface\cite{besson2015antarctic}. Anomalous enhancements (or suppressions) in surface reflectivity at a given locale are therefore evident from a measurement of the ratio of the reflected Solar RF signal strength to the direct Solar RF signal strength at that location. Since solar emissions are incoherent and broadband, possible sub-surface reflectors should be manifest as local enhancements in the measured solar albedo at a given location. Figures \ref{fig:A2VPolSolarPower} and \ref{fig:A2HPolSolarPower} compare the observed reflected power with the Fresnel expectation for vertically-polarized vs. horizontally-polarized signal, as a function of location on the Antarctic continent. Superimposed on the Figures are the locations of AE1 from the ANITA-1 mission and AE2 from the ANITA-3 mission. No obvious enhancements in observed reflected power are observed at the location of AE1. 


\subsection*{HiCal data} Complementary to the continuous emissions from the Sun, the balloon-borne HiCal transmitter is designed
to calibrate ANITA's response to surface reflections via a high-voltage ($\sim$2kV), short-duration ($\sim$10 ns) RF signal
emitted from a 35--38 km elevation, and horizontally displaced from ANITA by 100--1000 km. 
The HiCal-1 mission\cite{gorham2017antarctic} flew in tandem with ANITA-3, providing 100 `doublets' of RF signals observed both directly (`{\tt D}') and also
via surface reflections (`{\tt R}'); such pairs are readily identified by their predictable, and geometry-dependent {\cal O}(10) microsecond
time delay between pulses. 
HiCal-2 comprised two Dec., 2017 flights (``a'' and ``b'', launched in reverse chronological order)\cite{prohira2019hical}, both of which provided an order-of-magnitude improvement
in doublet statistics over HiCal-1.


\section*{Shallow Sub-surface reflectors ({\tt SSR})} 
\subsection*{Overview} Shoemaker {\it et al.} posit that the anomalous events may result from the presence of embedded layers with local
density `inversions', such as summer surface water pools which freeze and then compactify in
successive years. Indeed, such widespread surface pooling around the low-elevation Antarctic margins
has recently been deduced from digital processing of images taken with the
Landsat 8 and Sentinel 2A satellites\cite{stokes2019widespread}. 
Although such features may extend up to 500 km inland and also
at $>$1.5 km elevations, the total fractional aereal coverage of such features is, in any case, $<10^{-4}$.

Frozen-in ice layers having meter-scale thickness (comparable to radio frequency wavelengths) have not been reported in ice core data\cite{Fujita2006Z}.
Since the reflection efficiency for such an embedded layer (depending on the local dielectric contrast) rapidly decreases with depth, asymptotically approaching zero below the firn as the density of the ice layer approaches the density of the surrounding ice, we primarily consider near-surface layers here. At a typical interior surface snow accumulation rate of $\sim$10 cm/yr (characteristic of South Pole, e.g.), such a near-surface layer would presumably have to have been generated within the last 20-30 years. Non-ice layers, at all depths, and in both Greenland and Antarctica, have been extensively studied in radar surveys (CRESIS, BAS, UTIG, e.g.), 
although layer reflections rarely exceed -40 dB in return power and are therefore incapable of producing the significant broadband signals observed by ANITA.

\subsection*{Signal features}
Nanosecond-scale
RF pulses reflecting in proximity to 
 meter-scale sub-surface reflectors with meter-scale thickness $t$ and depth $d$ should result in multiple observable signals in a typical $\sim$100-ns ANITA waveform capture window. 
In sequence, the air/surface reflection arrives at the
payload earliest, with subsequent reflections due to the snow/ice interface, etc; each layer, of course, admits multiple
internal reflections before signal emerges back upwards towards the payload. 
After unfolding the system response, 
none of the observed ANITA cosmic ray events, however,
exhibit any obvious indication of such after-pulses\cite{HooverNamGorham2010,Ludwig2019radio,cao2018search}.
Shoemaker {\it et al.} evade the absence of after-pulses
by `tilting' the sub-surface reflector by an angle exceeding $\sim\delta\theta_C$, such that the sub-degree aperture geomagnetic signal surface
reflection
is outside the payload's acceptance. The tilt is set to the required inclination angle such that only the sub-surface reflection
is within ANITA's solid angle acceptance, resulting in only one observed pulse (multiple internal reflections within the layer are presumably too small in amplitude to be visible). Although no experimental evidence is presented in the Shoemaker {\it et al.} for such near-surface tilted layers, to fit their model, such layers would require local non-uniform (and linearly increasing/decreasing) snow overburden accumulation at a putative reflection site over a scale of hundreds of meters. Since the HiCal beam is much broader than the one-degree scale of UHECR emissions, signals emitted by HiCal should be observed in both their surface, as well as sub-surface reflections, for all geometries.

\section*{Model-dependent measures of sub-surface reflectors}
In what follows, we test the Shoemaker {\it et al.} model by building, from the HiCal/ANITA data itself, synthetic {\tt SSR} waveforms. We then compare our {\tt R} data waveforms to the simulated {\tt SSR} waveforms and also to the {\tt D} data waveforms. If sub-surface reflectors are present, they should correlate better to {\tt R} than {\tt D} waveforms correlate to {\tt R}. However, in the absence of any transmitter beam pattern anisotropies, and assuming that the surface reflection is specular, {\tt R} should be a reduced amplitude, inverted copy of {\tt D}.

\subsection*{Model 1: Embedded ice layers with meter-scale transverse thickness}
\begin{figure}[htpb]\centerline{\includegraphics[width=0.8\textwidth]{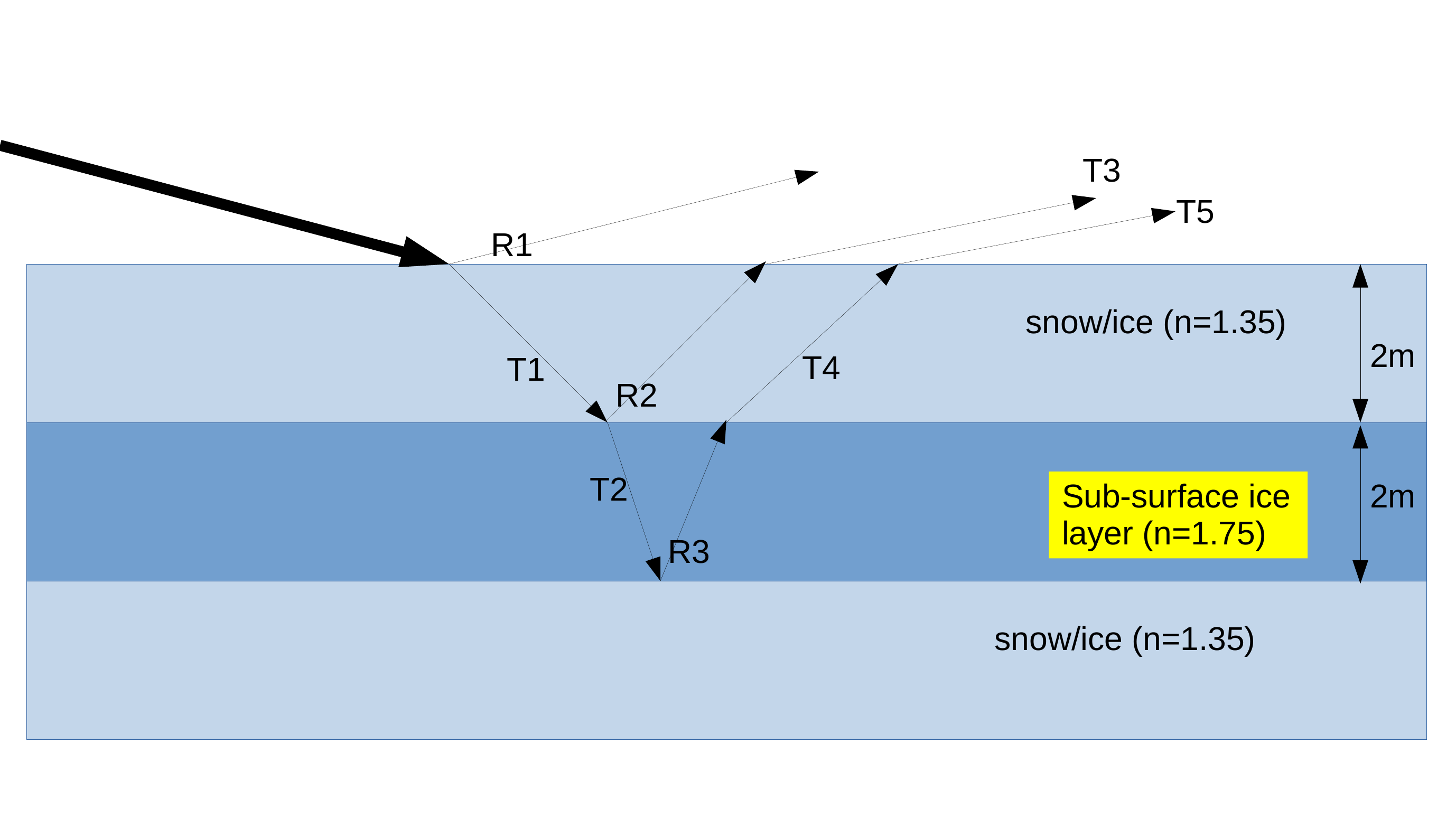}}\caption{Model 1 for embedded sub-surface ice layer, illustrating primary reflected ray paths. Each interface is characterized by a calculable reflection (``R'') or transmission (``T'') Fresnel coefficient. Most relevant for our case are the three signals shown emerging with an upwards trajectory (R1, T3 and T5, the latter of which corresponds to anomalous polarity).} \label{fig:rays}\end{figure}

We consider two {\tt SSR} models. The first, comprising a single meter-scale layer based on ice shelf measurements of 
embedded pond-melt pools with 170 $kg/m^3$ density contrasts found on the Larsen Ice Shelf\cite{hubbard2016massive} is
illustrated in Figure \ref{fig:rays}; similar features have also been found in Dronning Maud Land\cite{Liston2005antarctic} and
predicted elsewhere\cite{Lenaerts2016present,stokes2019widespread}.
Following Shoemaker {\it et al.}, we neglect secondary (tertiary, etc.) reflections from signal `trapped' within one layer.
We stress that this model is intended to be generic, yet representative of glaciological features
capable of producing the high-fidelity, ns-scale impulses
characteristic of the observed ANITA anomalous events.
Since typical ice cores show layering with centimeter scales and fractional density contrasts rarely exceeding 5\% (considered below as `Model 2'),
the features depicted in Fig. \ref{fig:rays} are therefore somewhat extreme,
imagining a solid ice (n=1.78) layer with thickness $t_i$ buried in snow a distance $t_s$ below the surface. Realistic local density contrasts are at least an order-of-magnitude smaller than the solid-ice layer assumption herein.

For simplicity, we take all surfaces to be parallel. We set $t_i$=$t_s$=2 m as nominal values and consistent with the 2-3 m values found in \cite{hubbard2016massive}; the relative amplitudes of observable signals depend only weakly on the choices of thickness and depend primarily on the incidence angles and density contrasts. For meter-scale layers, the relative time delays between observable signals will scale linearly with the chosen values of $t_s$ or $t_i$. At the $i^{th}$ interface, for a given incidence angle $\theta_i$ (and transmission angle $\theta_t$, given by Snell's Law), we calculate the Fresnel amplitude reflection coefficient $R_i$ and amplitude transmission coefficient $T_i$, with the energy-conservation constraint\footnote{Here the $\sin^2$ terms account for the focusing/de-focusing of signal flux as rays refract towards/away-from the normal across an interface.} that $R_i\sin^2\theta_i=T_i\sin^2\theta_t$. In practice, although the energy-conservation constraint affects the calculated in-ice signal strengths, the correction for signal penetrating into the ice is canceled by an inverse correction for signal reflected upwards and emerging from the ice.
The limited event buffer for the ANITA data acquisition system restricts ANITA's sensitivity, within a single waveform capture, to layers with a maximum total depth of no more than $\sim$4 meters. Assuming that ANITA triggers on the initial air-surface reflection, the reflection from the lower surface of ice layers deeper than 4 meters would typically appear at times beyond the upper edge of the single-event waveform buffer. 

\begin{figure}[htpb]\centerline{\includegraphics[width=0.8\textwidth]{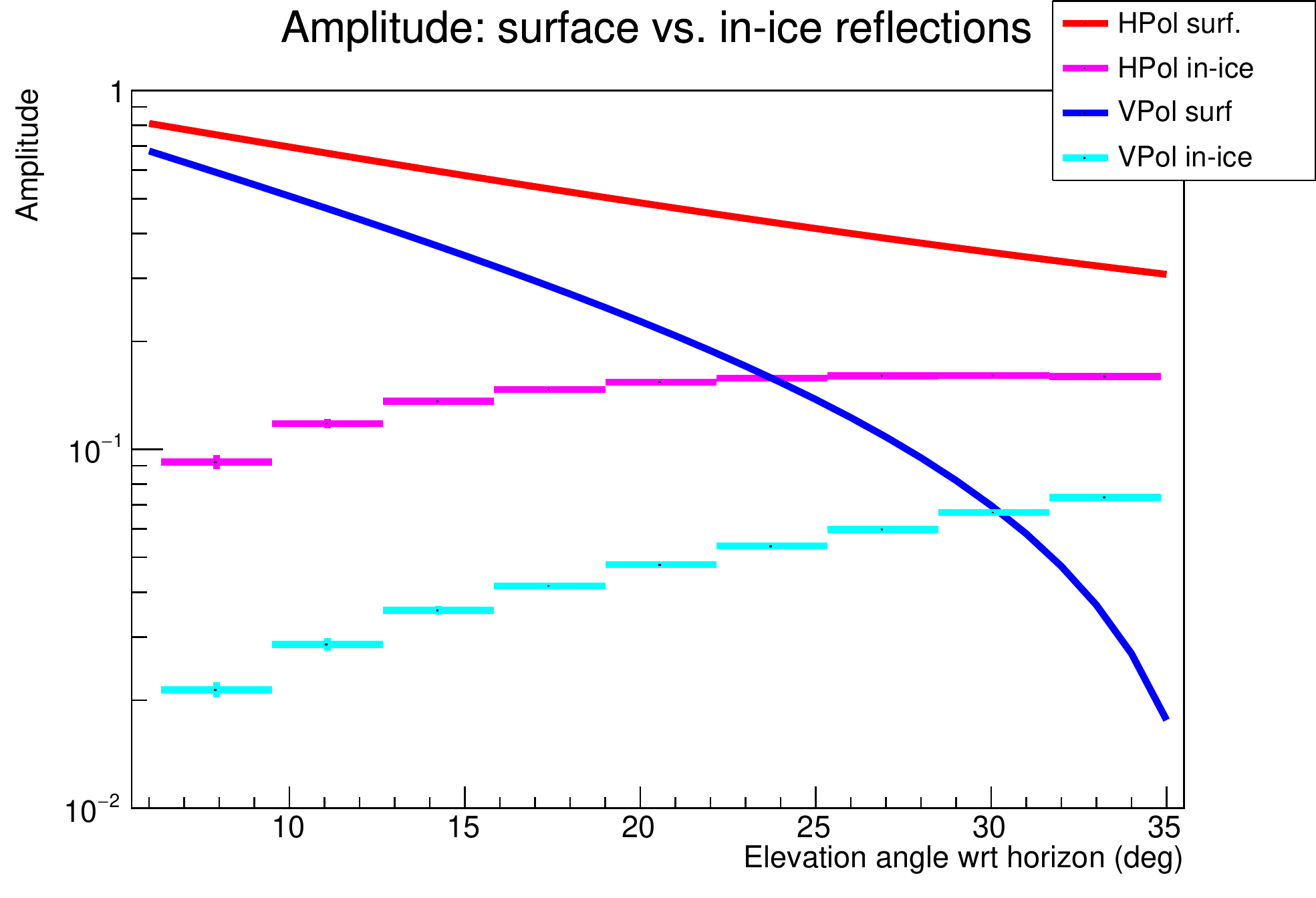}}\caption{Expected amplitude of direct surface reflection, relative to putative sub-surface, 2 meter thick in-ice layer, as a function of elevation angle.\label{fig:ReflModel1}}\end{figure}

Taking into account all relevant transmission and reflection coefficients,
Figure \ref{fig:ReflModel1} shows the relative amplitudes of the expected reflections, comparing T3 to R1 only.
From the Figure, it is immediately evident that the (dominant) HPol signal amplitude for a sub-surface reflector is considerably suppressed
relative to the initial surface reflection. Qualitatively, the Shoemaker {\it et al.} model requires a high-enough transmission coefficient for signal to reach the sub-surface layer, but this is at the expense of significant in-ice reflection. This can be ameliorated somewhat by 
extending the sub-surface layer fully to the surface, although such a variant is somewhat {\it ad hoc}.
Note that if we take a more realistic, but still quite large case of a density contrast for an embedded layer of 5\% rather than the 30\% assumed here, the magnitude of the anomalous polarity T5
 signal relative to R1 falls below 10\%, and therefore less than half of the retained-signal amplitude required by Shoemaker {\it et al.}. This again would imply a significantly higher mean energy cosmic ray primary observed via sub-surface reflections than is otherwise observed for ANITA's UHECR sample observed via surface reflections, and, as mentioned previously, difficult to accommodate given an $E^{-2.7}$ power law UHECR primary spectrum.

\begin{figure}[htpb]\includegraphics[width=0.8\textwidth]{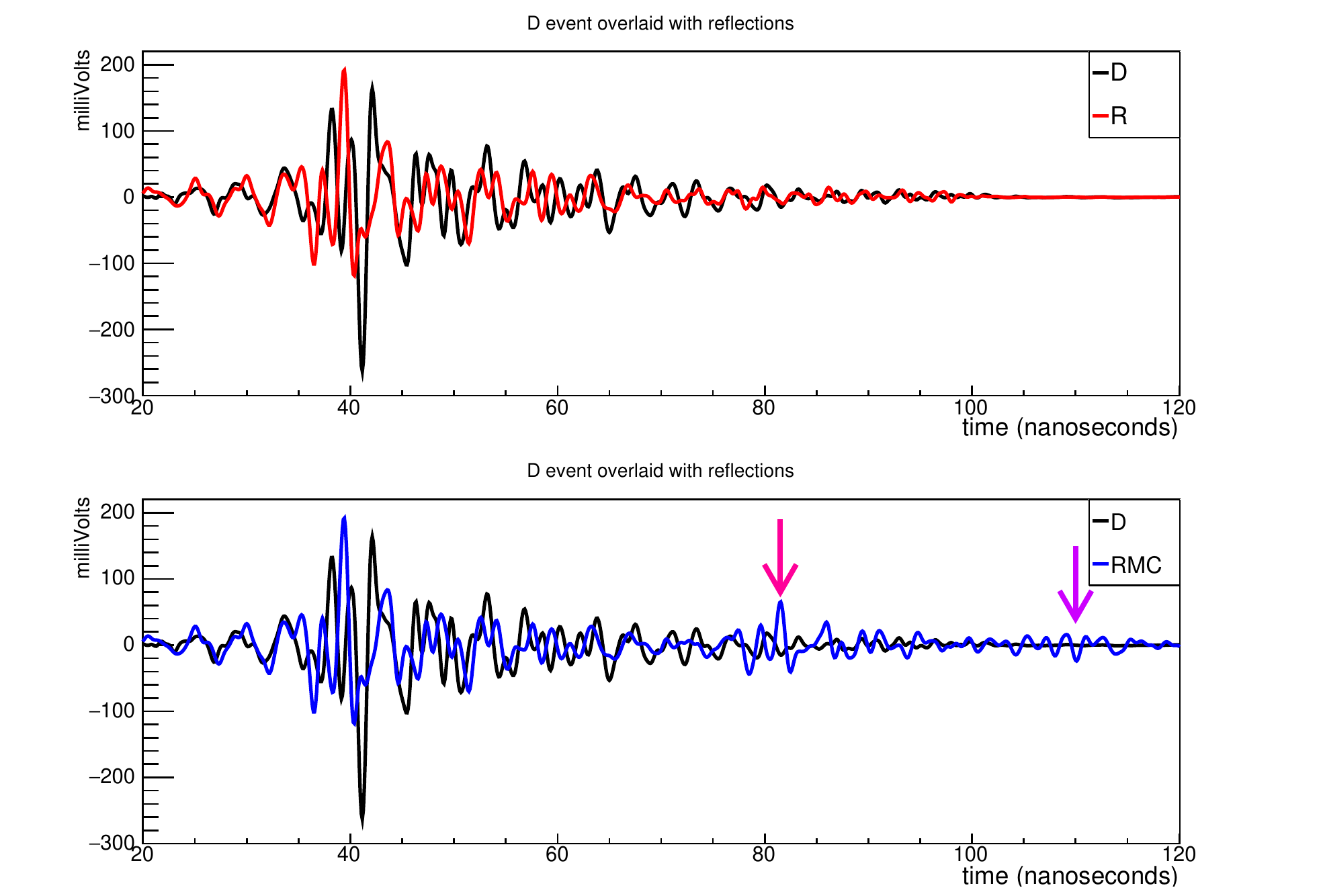}
\caption{HiCal ${\tt D}$ overlaid with ${\tt R}$ events (top) and (bottom) overlaid with simulated {\tt RMC} signals resulting
from sub-surface reflections, using procedure described in the text. Magenta and violet arrows show simulated reflections from top and bottom of embedded layer, respectively. The latter pulse has `anomalous' polarity.}\label{fig:hc1bev}\end{figure}

We have simulated the waveform resulting from an embedded, 2-meter thick ice layer two meters below the snow surface on the HiCal reflected signals observed by
ANITA. Subsequent to the initial air/surface-snow reflection, we expect an inverted secondary reflection (T3), 
delayed by $\sim$25 ns from the ice-top-layer/snow interface, followed by a tertiary signal (T5) from the lower ice-layer/snow interface
delayed by an addition $\sim$25 ns, with polarity opposite the first two observed signals.
In this way, we build synthetic Reflected Monte Carlo ANITA waveforms (``{\tt RMC}'') expected from sub-surface, englacial, meter-scale layers by adding to captured {\tt D}-events appropriate
`copies' of that same event, delayed and/or inverted, as appropriate for the ice-top-layer/snow and/or bottom-ice-layer/snow reflections, with amplitudes
determined from Snell's Law and the appropriate Fresnel coefficients.
If englacial embedded layers are prevalent in Antarctica, the observed reflected HiCal-2 {\tt R} waveforms 
should match better to our modeled {\tt RMC} waveforms
than the original {\tt D} waveforms.
A typical HiCal reflected event ({\tt R}), 
as recorded in ANITA-4, is presented in Figure \ref{fig:hc1bev}, displaying the coherently-summed
waveform
as captured by the ANITA Labrador digitizer corrected for the system response of the ANITA signal chain. 
Also shown in Figure \ref{fig:hc1bev} is the modeled {\tt RMC} signal form, with the secondary T3 and tertiary T5 amplitude normalizations as expected for this incidence angle.

\subsubsection*{Comparison to Model}
\begin{figure}[htpb]\includegraphics[width=0.8\textwidth]{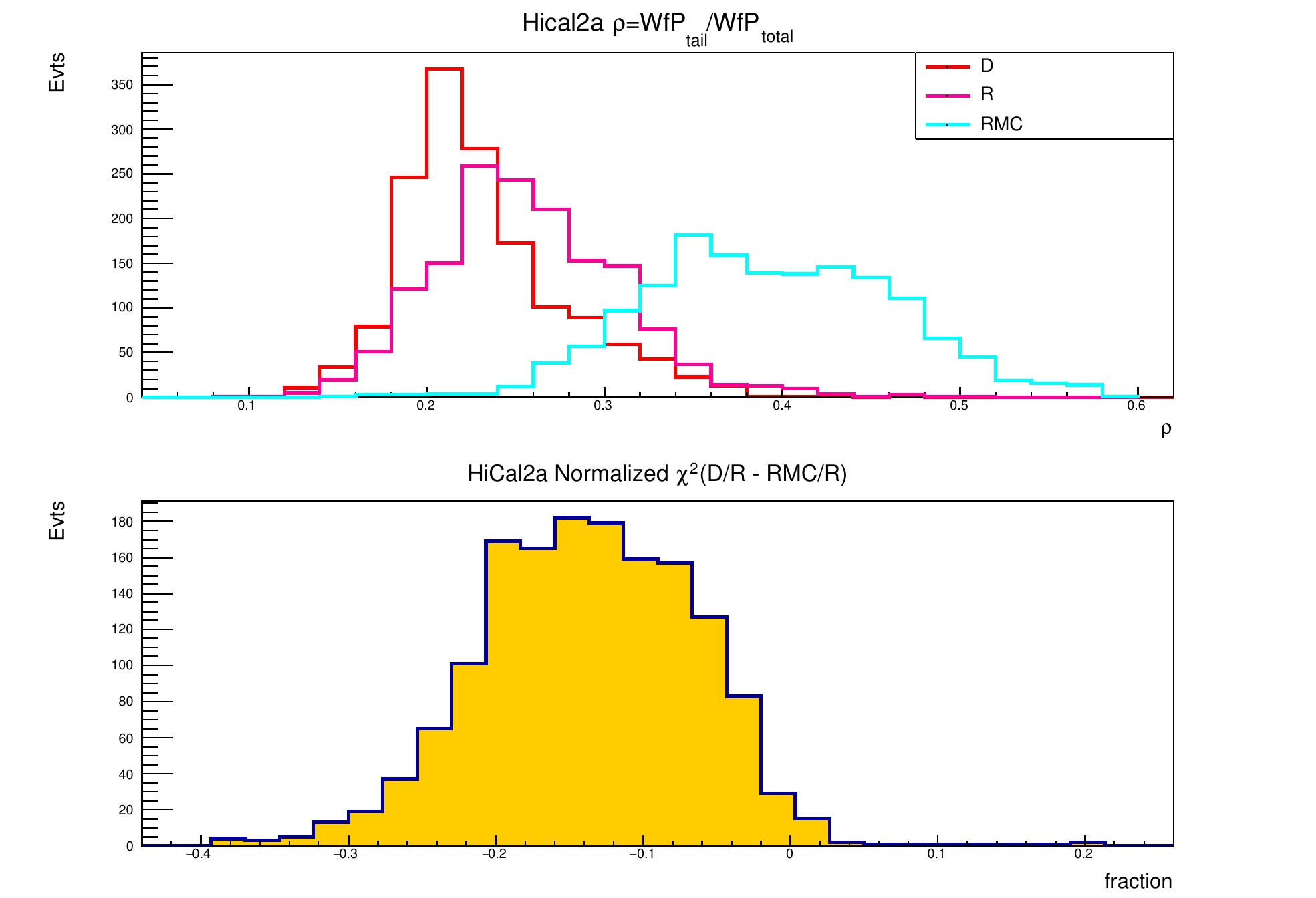}\caption{HiCal 2a: Top panel: ratio of signal waveform power for samples 20 ns beyond peak value, relative to total beyond-peak power in waveform for direct (``{\tt D}''), reflected (``{\tt R}'') and modeled embedded-layer reflected (``{\tt RMC}''). We note significant excess tail power for {\tt RMC} sample, as expected for after-pulses resulting from secondary (/tertiary) reflections. Bottom: Difference in $\chi^2$ between
{\tt R}/{\tt D} waveform shapes vs. 
{\tt R}/{\tt RMC} waveform shapes. Negative values imply preferable match of {\tt D} waveform to {\tt R} waveform.}\label{fig:R2RMC_WfP_hc2a}\end{figure}

\begin{figure}[htpb]\includegraphics[width=0.8\textwidth]{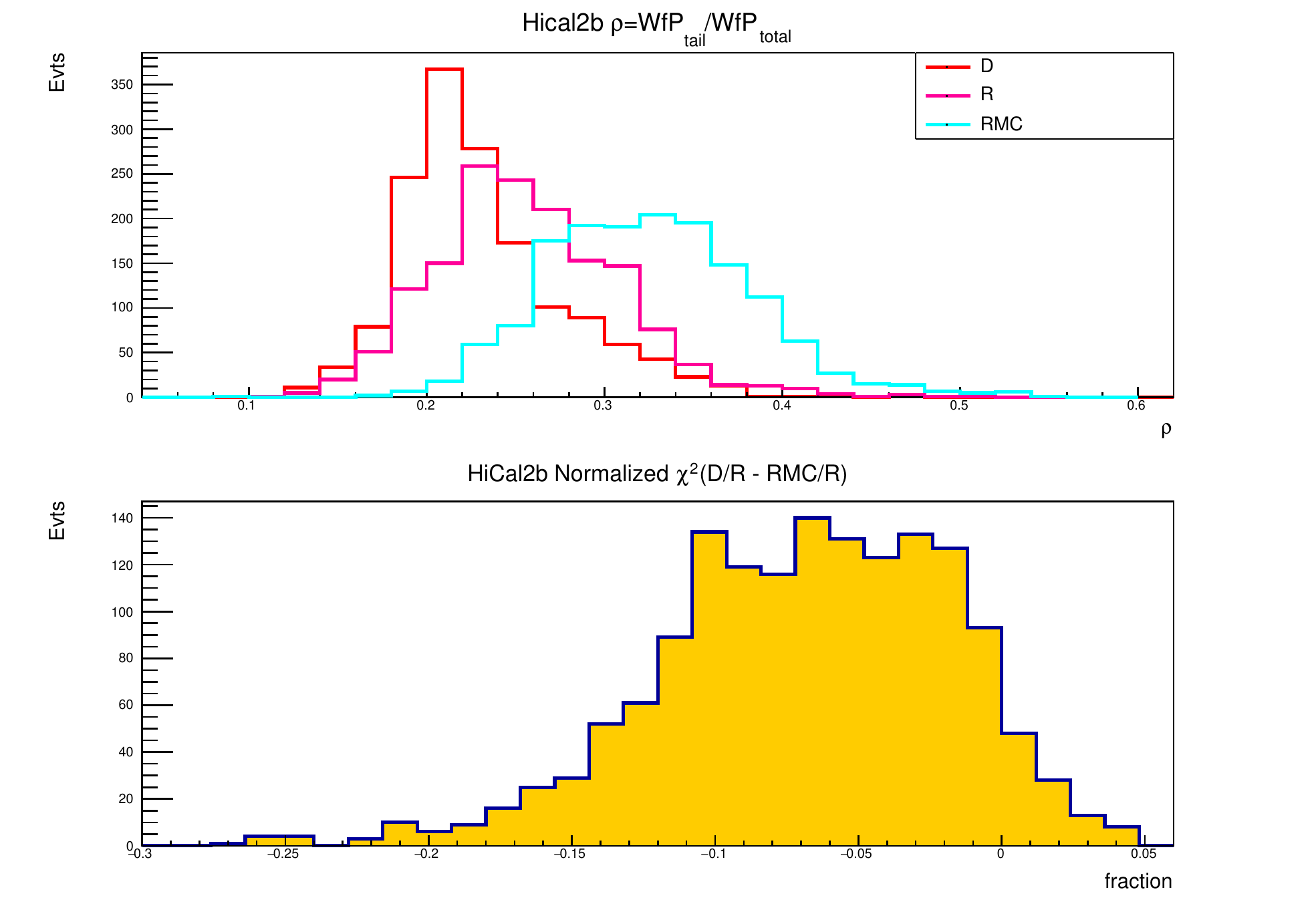}\caption{HiCal 2b: Top panel: 
ratio of signal waveform power for samples at least 20 ns beyond peak value, 
relative to total power for samples beyond peak value for {\tt D}, {\tt R}, and modeled embedded-layer reflected ({\tt RMC}). 
We note significantly more relative tail power for {\tt RMC} sample, as expected for after-pulses resulting from secondary (/tertiary) reflections. Bottom: Difference in $\chi^2$ (see definition in text) between 
{\tt R}/{\tt D} waveform shapes vs.
{\tt R}/{\tt RMC} waveform shapes. Negative values imply preferable match of D waveform to {\tt R} waveform.}\label{fig:R2RMC_WfP_hc2b}\end{figure}

To test the {\tt SSR} model, we use 
the ratio $\rho$ of the total waveform power in the `tail' (defined as at least 20 ns beyond the peak voltage in the waveform, so $\rho=\Sigma(V_i^2[t_i>t_{peak}+20~{\rm ns}])/\Sigma(V_i^2[t>t_{peak}])$, where
$t_{peak}$ is the time of the peak voltage in the waveform), 
relative to the total power of a captured waveform (beyond the peak sample) to
quantify the consistency of the HiCal reflected {\tt R} data sample 
with the embedded ice layer hypothesis. In Figures \ref{fig:R2RMC_WfP_hc2a} and \ref{fig:R2RMC_WfP_hc2b}, we compare the $\rho$ distribution for both {\tt R} vs. {\tt D} and also {\tt R} vs. {\tt RMC}. In each Figure, the bottom panel depicts the $\chi^2$ difference for {\tt R} compared to {\tt D} vs. {\tt R} compared to {\tt RMC}, with $\chi^2$ defined as ($\rho_{\tt D}/\rho_{\tt R}$ -$\rho_{\tt RMC}/\rho_{\tt R})^2$.
The fraction of events favoring the {\tt RMC} hypothesis in the HiCal-2a(/2b) data sample is 1.3\%/(3.6\%), somewhat smaller than Shoemaker {\it et al.}'s 7\% criterion. We note that this fraction from HiCal should be interpreted as an upper bound. Some HiCal events may have contamination from  anthropogenic RFI in a similar direction and which would spuriously increase the amount of power in the tails. Moreover, an explanation of AE1 and AE2 from this subsurface mechanism requires a minimum relative tilt over a wide area, which would only be present in a subset of events.

\begin{figure}[htpb]
\includegraphics[width=0.6\textwidth]{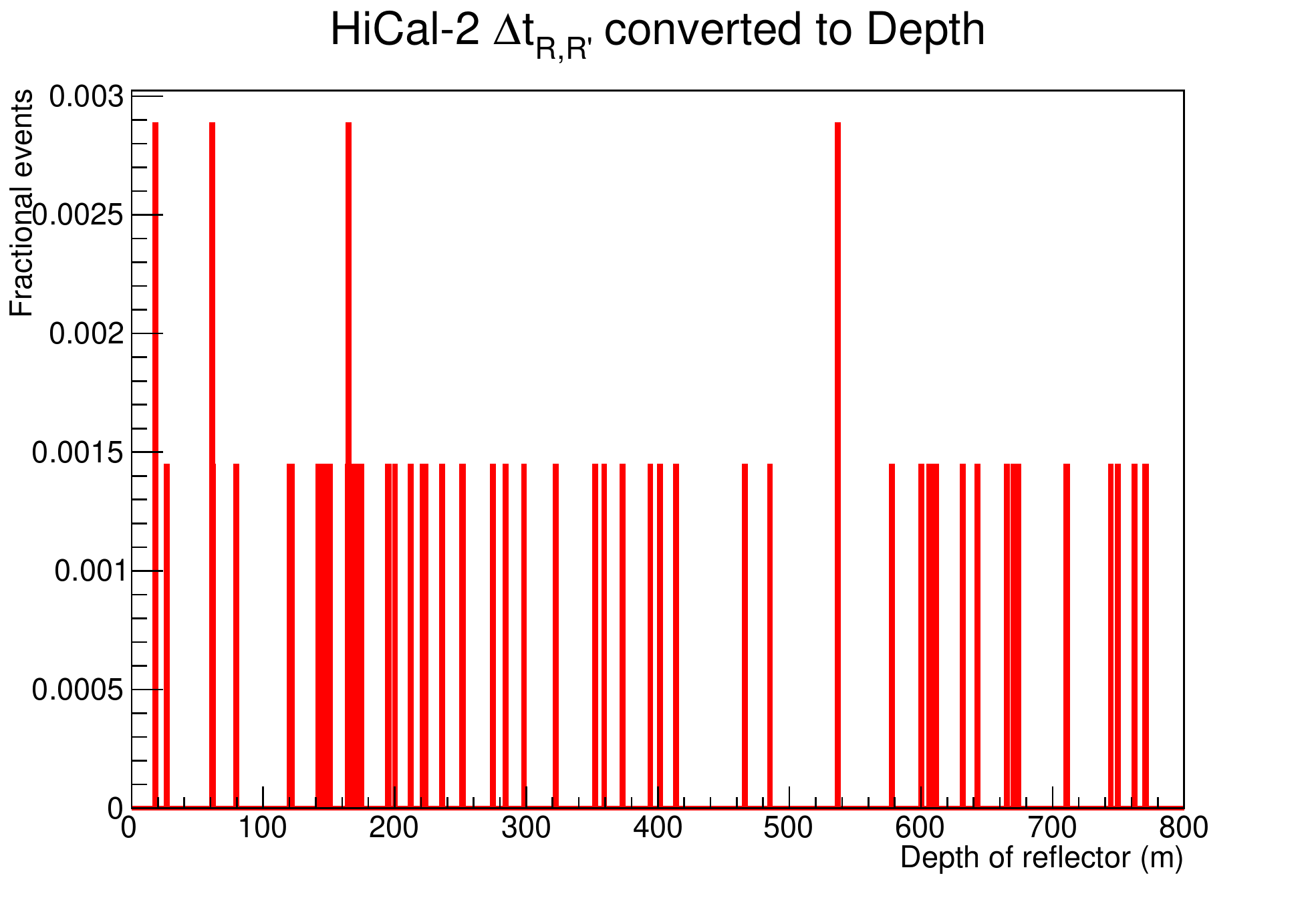}
\caption{Time difference between an {\tt R} event and the subsequent ANITA-4 trigger, converted into implied depth. An embedded layer at a fixed depth would be expected to produce some enhancement, above background, in this distribution.}
\label{fig:dtAiAjSSR}
\end{figure}

Embedded reflectors deeper than 4-5 meters will not be registered in the same ANITA event trigger, but could, if the signal strength is large enough, produce triggers in successive events. Figure \ref{fig:dtAiAjSSR} shows the HiCal-2a time difference between an {\tt R} event and the subsequent trigger registered by ANITA, converted to depth. At a nominal 50 Hz trigger rate, the typical time-between-triggers is 1/50 Hz, or 2000 microseconds. For an embedded reflector, a second trigger following {\tt R} would occur with an approximate delay of one microsecond (divided by the cosine of the angle with respect to the normal) for every 100 meters of layer depth. Again, no obvious depth clustering is observed in the Figure. (Unfortunately, rapid after-pulsing by the piezo-electric used to generate the RF pulse for HiCal-2b rules out a similar exercise using HiCal-2b pulser events.) 

\subsection*{Model 2: Multiple Internal Layers:}  
Internal layers, including thin yearly `crusts' due to summer surface melt and subsequent refreeze, as well as $<mm$-scale acidic conductive layers (resulting, e.g., from episodic volcanic activity) are common in both Antarctica and Greenland\cite{Fujita2006Z}.
For wavelengths comparable to the ray path between successive layers, interference maxima may be observed at appropriate viewing angles. The magnitude of such effects has been estimated using 
Finite-Difference Time Domain (FDTD) calculations. 
In principle, the spherical-wave decomposition (SWD) formalism developed to describe surface reflectivity as measured by the HiCal-2 experiment can also be used\cite{prohira2018antarctic} to quantify such effects.


We utilize the open-source MEEP software package for FDTD simulations. The simulation is composed of an electric field signal incident upon an ice reflector. The initial electric field is a plane wave delta function low-passed at 750 MHz incident at $60^\circ$ onto 15 m of ice. Following the specifications for internal reflectors given by Shoemaker {\it et al.}, the ice is modeled as multiple layers of dielectric with thicknesses ranging from 3 to 15 cm and with indices of refraction alternating between $n = 1.3$ and $n = 1.6$. This thickness range matches typical yearly snow accumulations, and the selected refractive indices correspond to surface-melt refreeze in alternating years. 
The azimuthal symmetry of ice layers over distances larger than one Fresnel zone in the radio `light pool' produced by a cosmic ray interaction allows our FDTD simulations to be restricted to 2-dimensions with no loss of generality.

\begin{figure}[htpb]\includegraphics[width=0.65\textwidth]{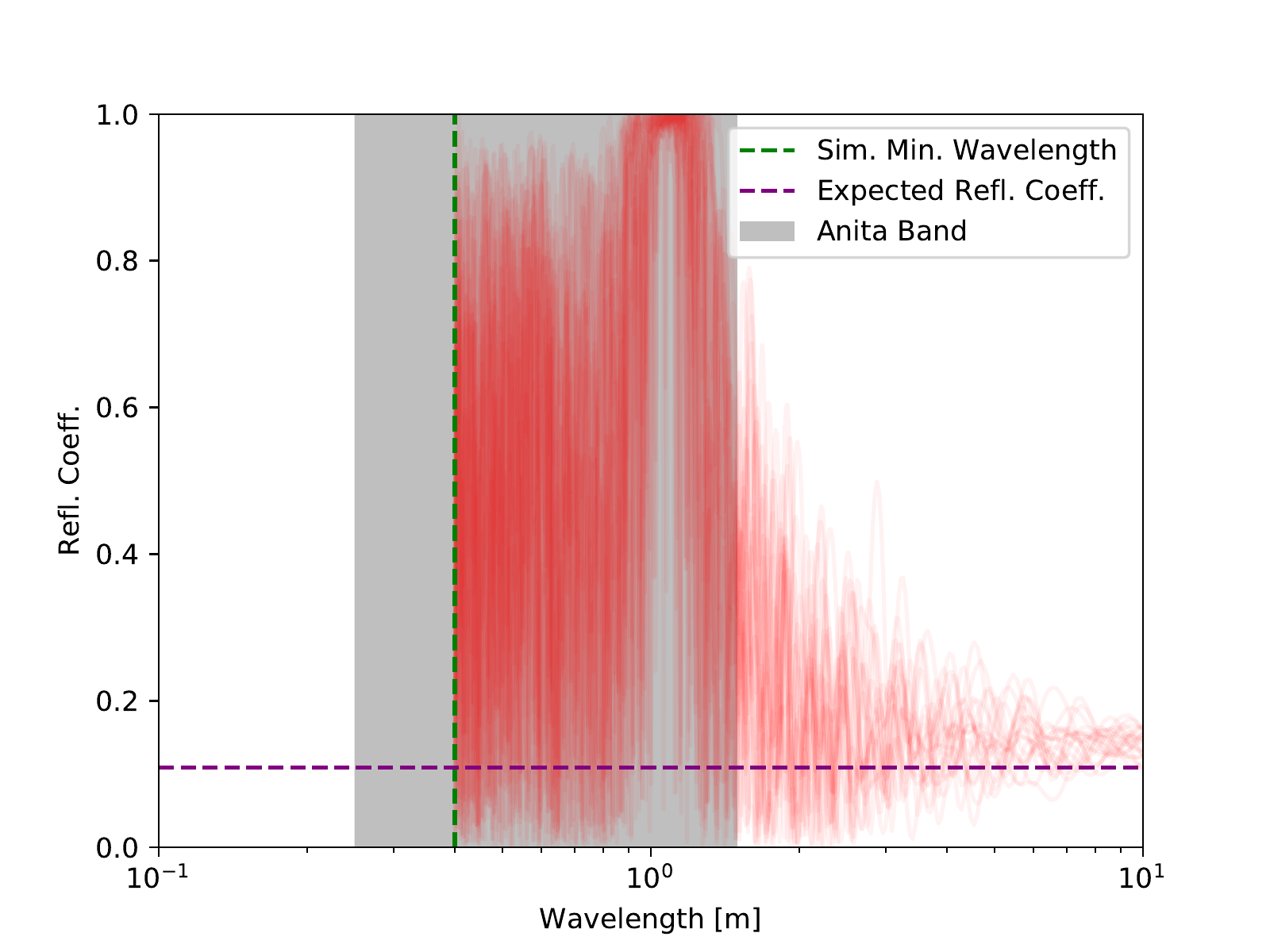}\caption{Reflection coefficients in power for scattering from the multiple thin-layer internal {\tt SSR} model.}\label{fig:dana}\end{figure}

\begin{figure}[htpb]\includegraphics[width=0.65\textwidth]{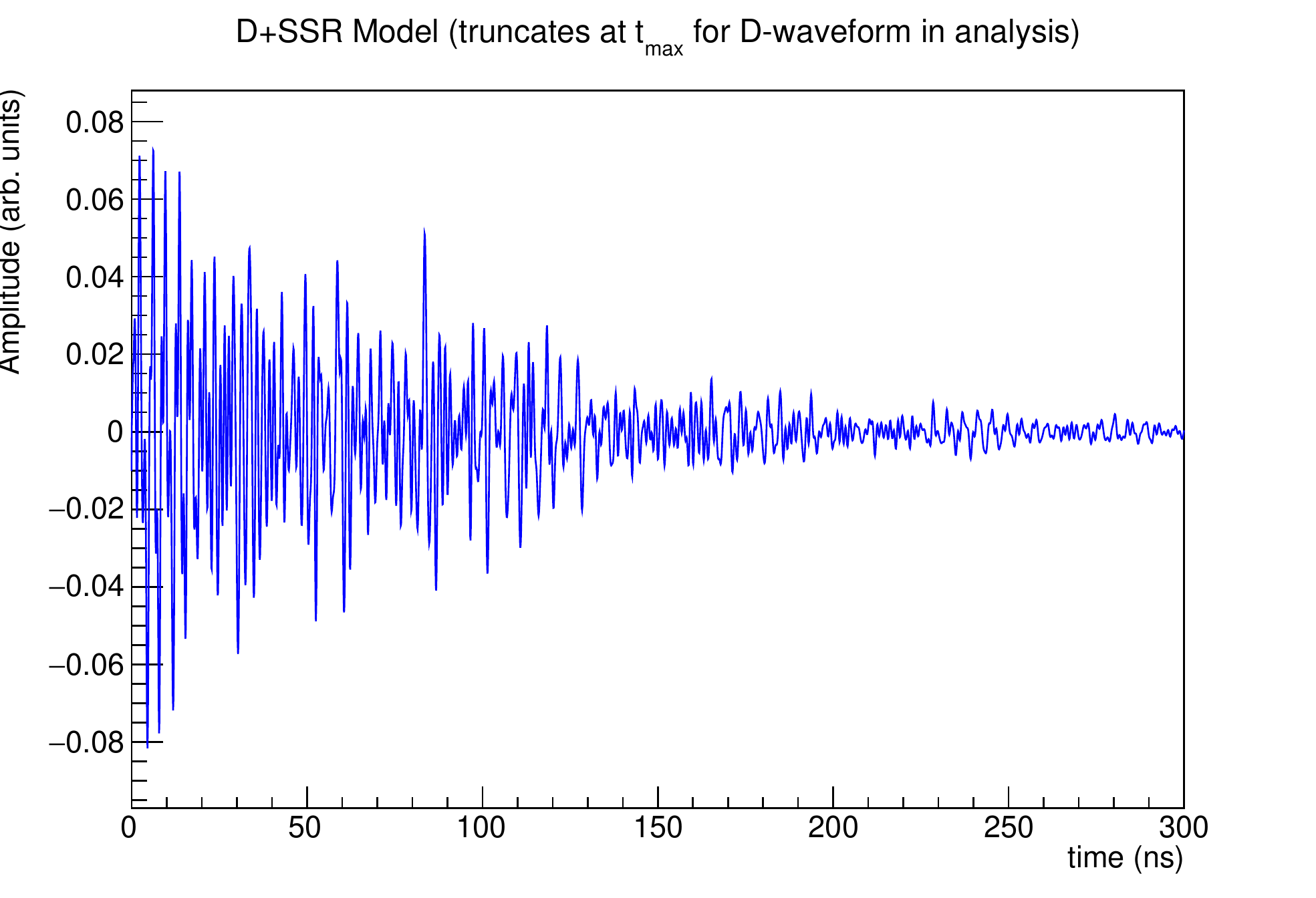}\caption{Simulated {\tt SSR} model-2 HiCal waveform. Note that the ANITA waveform capture is only of order $\sim$100  ns, some of which is pre-trigger time.}\label{fig:dangr}\end{figure}

After reflection from the ice surface, we calculate the net, far-field electric field using the near-to-far Green’s function. 
This approach leaves an ambiguous phase offset in the far-field signal. We correct for this unknown using the phase offset as derived from the far-field signal of the specular reflection of the incident signal. The final signal in the far-field is effectively the transfer function of the ice reflection, given the delta function input. 
The reflection coefficient from 20 iterations of randomized ice is plotted in Figure \ref{fig:dana}. While the reflection coefficient is indeed high for certain wavelengths in the ANITA band, this coefficient integrates over all times, while ANITA only records a relatively short window of around 100 ns. We approximate the ANITA trigger window by placing $t_0$ at $\sim20$ ns and truncating the signal at 128 ns. An example waveform for a single transfer response applied to the HiCal data is shown in Figure \ref{fig:dangr}. 

\begin{figure}[htpb]\includegraphics[width=0.65\textwidth]{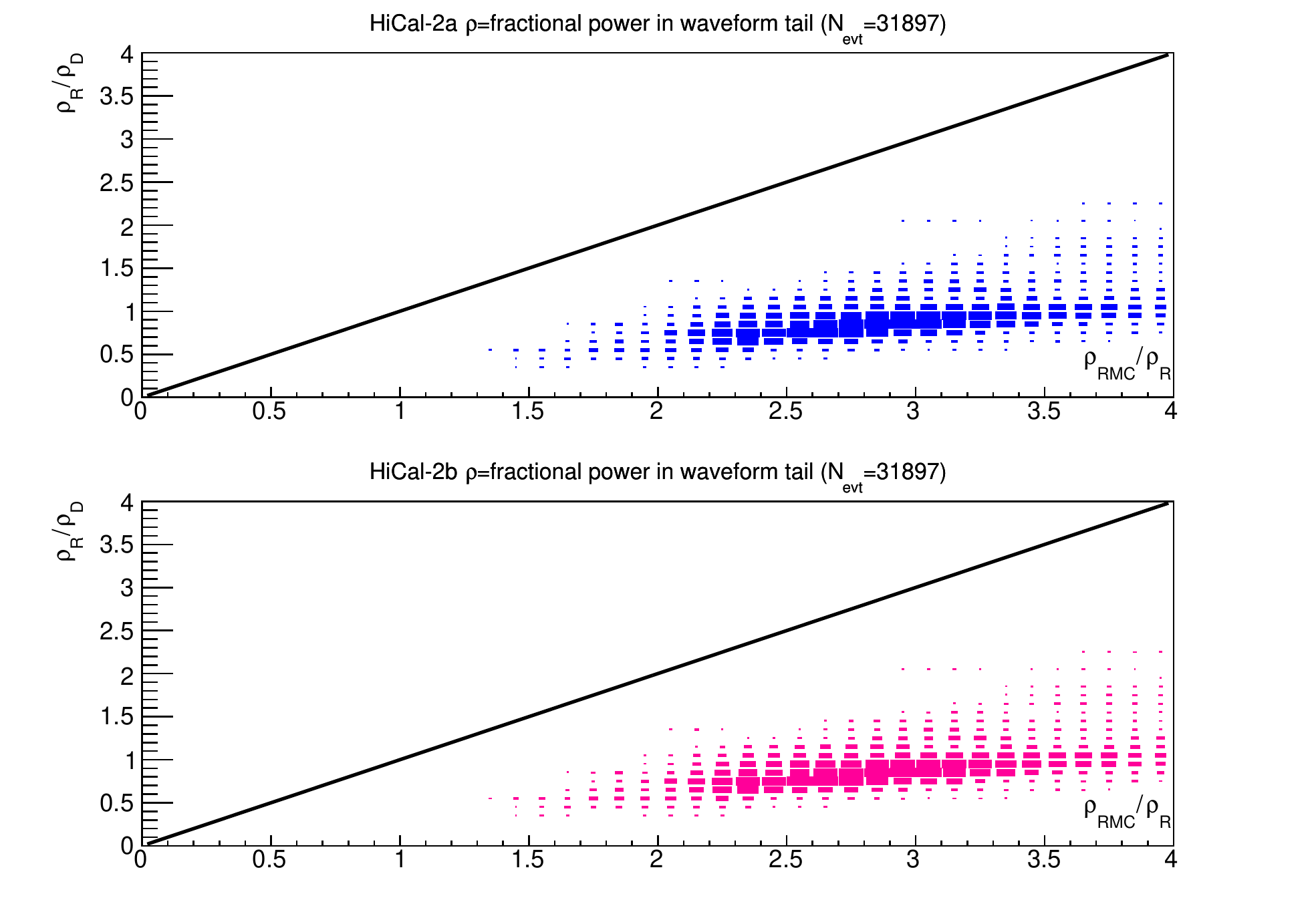} \caption{Comparison of HiCal-2 reflected ({\tt R}) data events to data ({\tt D}) vs. sub-surface reflector ({\tt SSR}) model. Above and to the left of the diagonal black line, {\tt SSR} model-2 is favored. Below and to the right of the black line, {\tt SSR} model-2 is disfavored. }\label{fig:WfPSSR}\end{figure}

Since the FDTD simulations indicate that {\tt SSR} should result in reflected signals with considerably
extended tails, we quantify our results using the previously defined parameter $\rho$. For specular reflection the {\tt R} waveform should be (modulo beam pattern effects) a reproduction of the {\tt D} waveform, such that there should be
the same fractional power in the tail for both {\tt D} and {\tt R}.
Figure \ref{fig:WfPSSR} compares the $\rho$ distribution for HiCal-2a data (top) vs. HiCal-2b data (bottom). In both cases, we note that the {\tt R} $\rho$ distribution, normalized to {\tt D}, cluster around a value of 1.0, consistent with the naive expectation that the observed R waveform should be a reproduction of the observed D waveform. By contrast, the {\tt SSR} $\rho$ distribution, normalized to {\tt R} events consistently exceed 1.0, consistent with the expectation that such reflectors should produce signal power received over {\cal O}(100 ns) rather than {\cal O}(10 ns). In no cases do we observe a preference for sub-surface reflectors over a single surface reflection in HiCal. Qualitatively, the long duration of the reflection from multiple layers is inconsistent with the comparatively short waveforms in AE1 and AE2. 

\section*{Conclusions}
Following Shoemaker $et~al$, we have considered several glaciological hypotheses offered to explain ANITA's `anomalous-polarity' AE1 and AE2 events. In several cases, the proferred models fail to produce adequate signal amplitude to match observation. Such cases can be accommodated only by stipulating that the primary UHECR causing the anomalous polarity reflections are approximately one-order magnitude higher in energy, causing significant tension with existing flux constraints. 
Satellite data, as well as ANITA data published within the last decade also do not readily indicate 
reflectivity anomalies at the Antarctic surface locations identified for AE1 and AE2. 
Using parameters specified by Shoemaker {\it et al.}, we have run simulations to test the hypothesis that sub-surface
reflectors are responsible for the observed anomalous events.
We find that the HiCal and ANITA data strongly disfavor
the Shoemaker $et~al.$ models.
Quantitatively, a model based on an embedded, tilted, meter-scale thickness near-surface ice-layer is allowed at the 2-3\% level; a model based on the coherent sum of reflections in thin, near-surface layers found over 7\% of the Antarctic surface is clearly disfavored. The recent report of four additional mystery events from the ANITA-4 mission at near-glancing angles, with 3.2$\sigma$ significance\cite{ANITA4ME}, is impossible to reconcile with the Shoemaker {\it et al.} model given the fact that $>$95\% of the incident signal amplitude is reflected at the surface for such a geometry, and would therefore be unable to penetrate to an embedded reflector, in any case.

\section*{Acknowledgments} 
ANITA-IV was supported through NASA grant
NNX15AC24G. We are especially grateful to the staff of
the Columbia Scientific Balloon Facility for their generous support. We would like to thank NASA, the National
Science Foundation, and those who dedicate their careers to making our science possible in Antarctica. This
work was supported by the Kavli Institute for Cosmological Physics at the University of Chicago. Computing resources were provided by the Research Computing Center at the University of Chicago and the Ohio
Supercomputing Center at The Ohio State University.
A. Connolly would like to thank the National Science
Foundation for their support through CAREER award
1255557. O. Banerjee and L. Cremonesi’s work was supported by collaborative visits funded by the Cosmology
and Astroparticle Student and Postdoc Exchange Network (CASPEN). The University College London group
was also supported by the Leverhulme Trust. The National Taiwan University group is supported by Taiwan’s
Ministry of Science and Technology (MOST) under its
Vanguard Program 106-2119-M-002-011.
 D. Besson and A. Novikov acknowledge the support from the MEPhI Academic Excellence Project (Contract No. 02.a03.21.0005) and the Megagrant 2013 program of Russia, via agreement 14.12.31.0006 from 24.06.2013. R. Nichol thanks the Leverhulme Trust for their support.


\bibliographystyle{unsrt} 

\end{document}